\documentclass[10pt,letterpaper,twocolumn]{article} 

\usepackage[tablesfirst,notablist,nomarkers]{endfloat} 

\usepackage{ol2}
\usepackage[draft]{hyperref}
\usepackage{amsmath}

\begin{document}

\twocolumn[ 

\title{Slow light of an amplitude modulated Gaussian pulse in electromagnetically induced transparency medium}


\author{Wenzhuo Tang, Bin Luo, Yu Liu, and Hong Guo$^*$}

\address{CREAM Group, State Key Laboratory of Advanced Optical
Communication Systems and Networks (Peking University) \\
and Institute of Quantum Electronics, EECS, Peking University, Beijing 100871, China \\
$^*$Corresponding author: hongguo@pku.edu.cn}

\begin{abstract}The slow light effects of an amplitude modulated
Gaussian (AMG) pulse in a cesium atomic vapor are presented. In a
single-$\Lambda$ type electromagnetically induced transparency (EIT)
medium, more severe distortion is observed for an AMG pulse than a
Gaussian one. Using Fourier spectrum analysis, we find that the
distortion, as well as the loss, is dominantly caused by linear
absorption than dispersion. Accordingly, a compensation method is
proposed to reshape the slow light pulse based on the transmission
spectrum. In addition, we find a novel way to obtain simultaneous
slow and fast light.\end{abstract}


] 

\noindent Recently, slow light has attracted tremendous attention
for its potential applications in optical communication. Group
velocity of light pulse can be slowed down using various effects,
such as electromagnetically induced transparency (EIT)
\cite{Harris1997,F-RMP}, double absorption line
\cite{Howell1,Howell2}, coherent population oscillations
\cite{cpo1,cpo2}, stimulated Brillouin scattering \cite{sbs1} and
stimulated Raman scattering \cite{srs1}, etc. Among these effects,
EIT is the most popular candidate to experimentally realize slow
light for, e.g., non-modulated Gaussian pulse. However, in real
communication, pulse modulation is of great importance, but has not
been well studied in slow light system.

In this Letter, we demonstrate the slow light effects for amplitude
modulated Gaussian (AMG) pulses in experiment, and analyze the
significant distortion of the AMG pulse. Further more, a
compensation method is proposed and a well-reshaped AMG pulse is
obtained.

\begin{figure}[htb]
\centerline{\includegraphics[width=8.3cm]{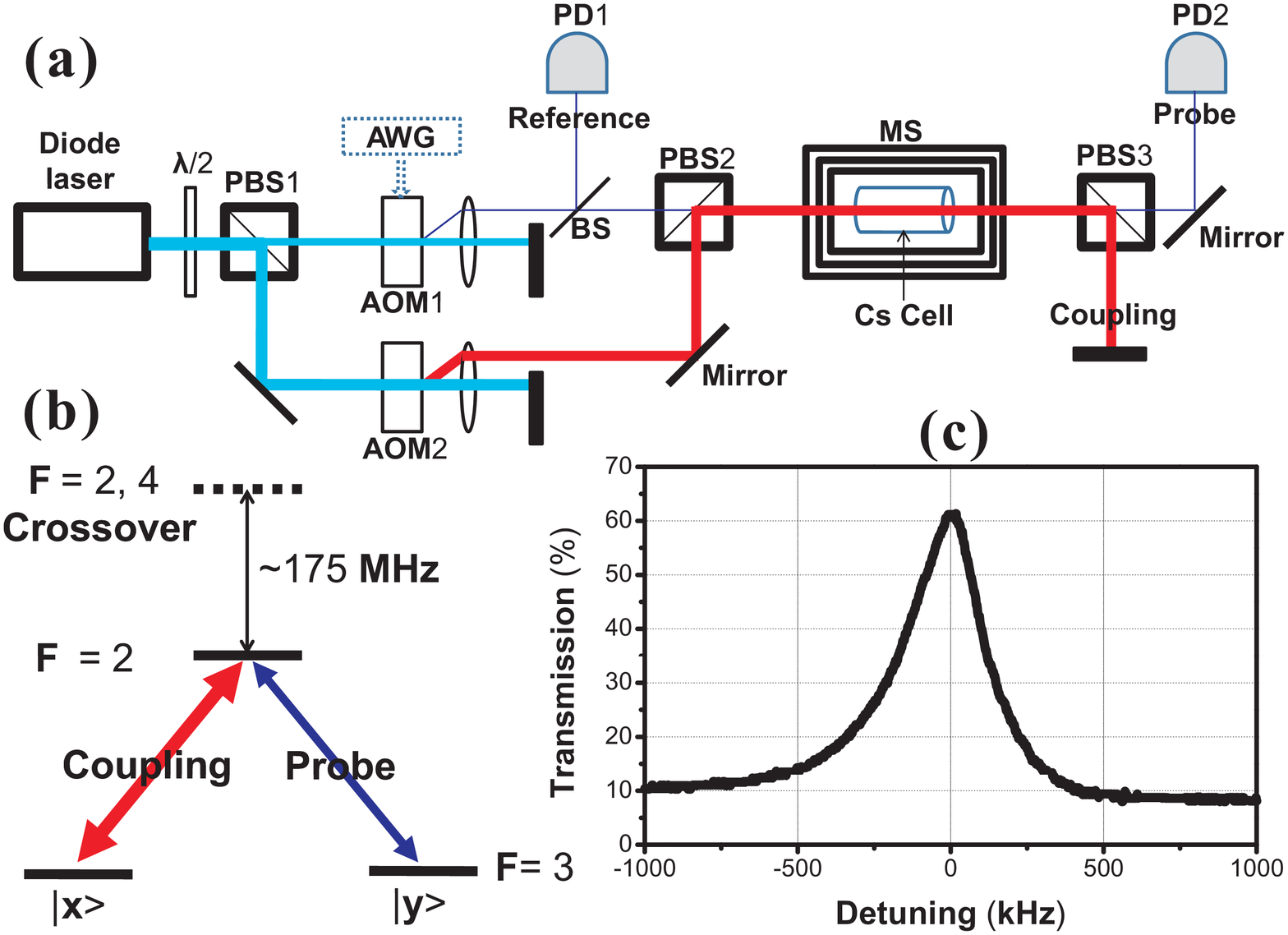}}
\caption{(color online). (a) Experimental setup. $\lambda /2$,
half-wave plate; PBS, polarizing beam splitter; AOM, acousto-optic
modulator; AWG, arbitrary waveform generator; MS, magnetic shield;
APD, \mbox{1 GHz} avalanche photo diode. (b) A simplified
single-$\Lambda$ type system of \mbox{$^{133}$Cs} atoms interacts
with the coupling and the probe lasers. Two ground states
$|x\rangle$ and $|y\rangle$ represent appropriate superpositions of
magnetic sublevels. (c) Probe transmission spectrum (an average of
16 measured results) versus detuning from resonance, with the
bandwidth of \mbox{$\sim$350 kHz} (FWHM) and the maximum of
\mbox{$\sim$61.5\%}.}\label{SETUP}
\end{figure}

Our experimental setup is schematically illustrated in Fig.
\ref{SETUP}(a). A homemade narrowband (\mbox{$\sim$300 kHz}) diode
laser is stabilized to the $D_2$ transition (852 nm) of $^{133}$Cs
atom: $F=3 \rightarrow F'=2, 4$ crossover. By the combination of
$\lambda/2$ (half-wave plate) and PBS1, the laser light is divided
into two beams with perpendicular polarizations and adjustable
proportion. Then, we can precisely control the optical frequency and
amplitude (thus intensity) of the coupling and the probe lasers
separately. The coupling laser frequency is redshifted
\mbox{$\sim$175 MHz} by AOM2, with power \mbox{$\sim$2.35 mW} and
$1/e^2$ beam diameter \mbox{$\sim$2.0 mm}. The probe pulse is
generated by passing a weak (\mbox{$<$10 $\mu$W}) continuous-wave
laser through AOM1, with the frequency redshifted \mbox{$\sim$175
MHz} and $1/e^2$ beam diameter \mbox{$\sim$2.0 mm}. In experiment,
the probe field amplitude $E(t)$ ($\geq 0$) is proportional to the
amplitude of the driving radio-frequency wave to AOM1, which is
directly controlled by an arbitrary waveform generator (AWG) [see
Fig. \ref{SETUP}(a)]. After that, both the coupling (vertically
polarized) and the probe (horizontally polarized) light are on
resonance with the $D_2$ transition $F=3 \rightarrow F'=2$ of
$^{133}$Cs atom [see Fig. \ref{SETUP}(b)], and are combined at PBS2.
Then, the overlapped lasers copropagate through the cell in order to
reduce the total Doppler width of the two-photon process
\cite{xiaomin1995}. The beam splitter (BS) in front of the cell
splits an appropriate portion of the probe to PD1 as reference,
whose intensity is set to be equal to that of output when the probe
is detuned far off-resonance. Thus, the background absorption is
eliminated and the reference can be treated as ``input". A \mbox{10
cm}-long paraffin-coated cesium vapor cell without buffer gas is
used at room temperature (\mbox{$\sim$25 $^{\circ}$C}) and is placed
in the magnetic shield (MS) to screen out the earth magnetic field.
The exit beams are separated by PBS3, so that only the slowed probe
beam reaches PD2 as the output.

\begin{figure}[htb]
\centerline{\includegraphics[width=8.0cm]{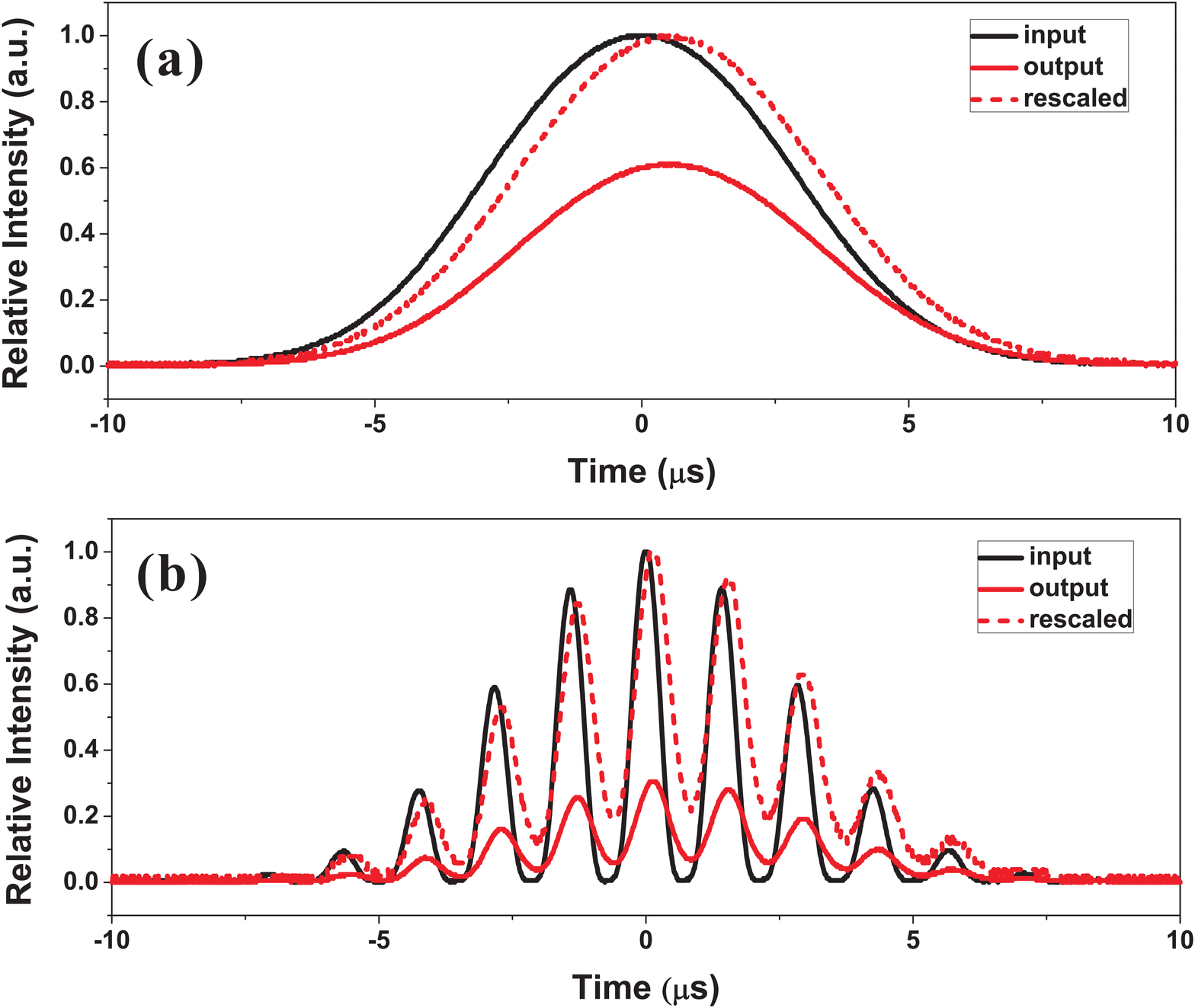}} \caption{(color
online). Input (normalized) and output pulses versus time, each
pulse is an average of 16 measured results; rescaled is the
normalized output. (a) Gaussian: low loss and little distortion. (b)
AMG: high loss and significant distortion. }\label{TIME}
\end{figure}

In experiment, the optical intensities $I(t)$ of the input and
output pulses are recorded on a \mbox{100 MHz} digital oscilloscope
triggered by an AWG. Thus, the amplitudes of both the input and
output (slowed) pulses can be obtained from the relation
$E(t)=\sqrt{I(t)}$. The two probe (input) pulses with nonnegative
amplitudes in time domain are chosen as
\begin{align}
\mathrm{Gaussian\ pulse:\ } &I_1(t)=\exp\left[-\frac{(\ln2)t^2}{T_0^2}\right],\nonumber \\
\mathrm{AMG\ pulse:\ }
&I_2(t)=I_1(t)\left[1+\cos(2\pi{\delta}t)\right]^2,\nonumber
\end{align}
where $I_1(t)$ and $I_2(t)$ are the intensities of two pulses,
respectively; \mbox{$\delta$ = 700 kHz} is the modulation frequency;
$T_0$ is the FWHM of the Gaussian pulse.

The experimental results are shown in Fig. \ref{TIME}. The two
pulses experience quite different slow light effects. For Gaussian
pulse case [Fig. \ref{TIME}(a)], it is delayed \mbox{$\sim$0.47
$\mu$s}, with relatively low loss and little distortion. Hence, we
can compensate for the loss by directly amplifying the pulse
intensity. However, for AMG pulse case [Fig. \ref{TIME}(b)], the
output undergoes relatively high loss and significant distortion,
while the delay time (\mbox{$\sim$0.14 $\mu$s}) is decreased. Thus,
the direct amplification is not feasible to compensate for its loss
and distortion [see the ``rescaled" pulse in Fig. \ref{TIME}(b)].

\begin{figure}[htb]
\centerline{\includegraphics[width=8.0cm]{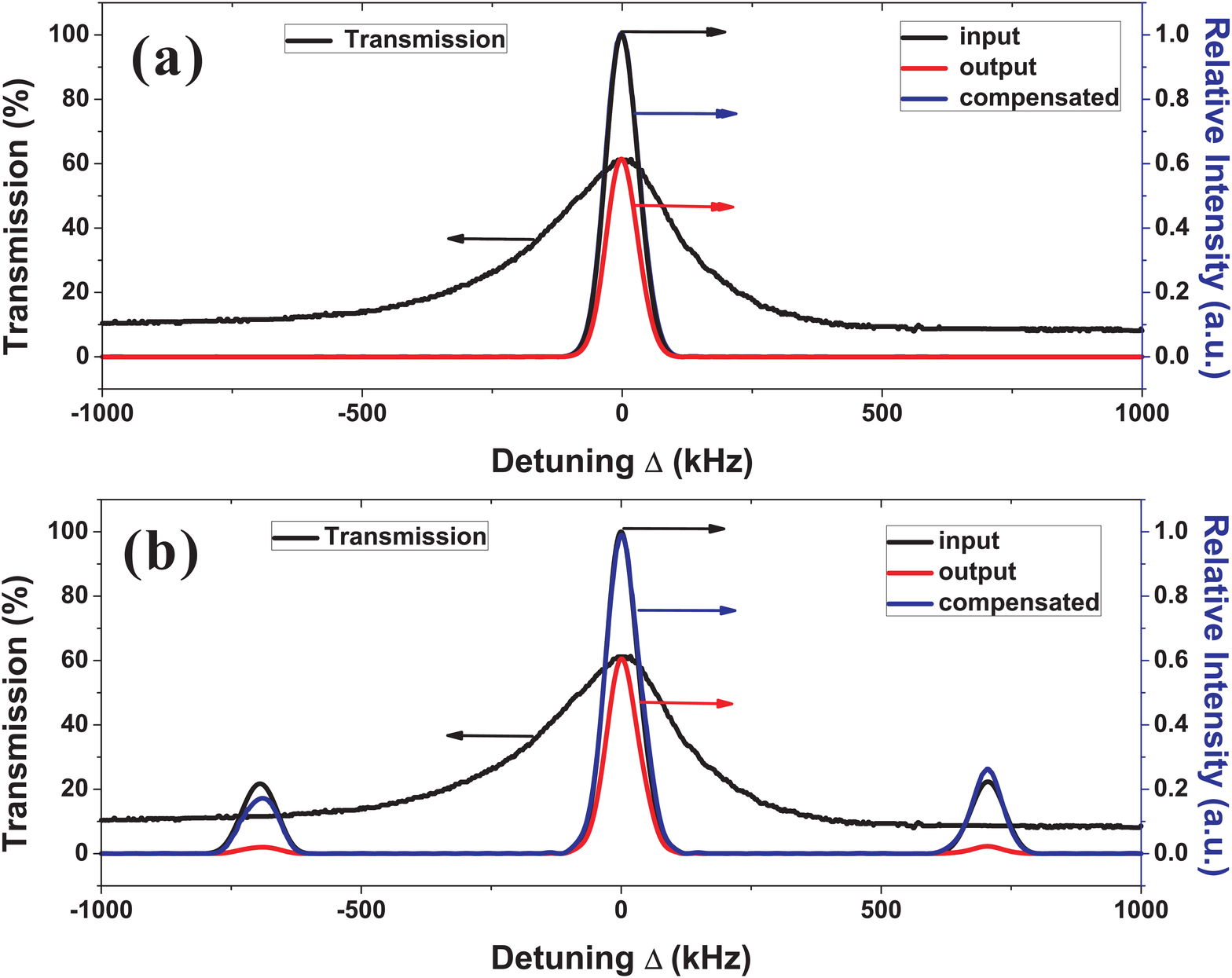}}
\caption{(color online). The intensity spectrums of input, output
and compensated pulse, and transmission spectrum. (a) Gaussian: the
compensated spectrum overlaps with the input one. (b) AMG: The
compensated components overlap with the corresponding input ones,
except that the sidebands have some minor deviations due to
experimental errors.}\label{FREQUENCY}
\end{figure}

It is known that the pulse distortion is caused by both absorption
and dispersion \cite{Michael2005}. We first analyze the ``absorptive
distortion" using the transmission spectrum [see Fig.
\ref{SETUP}(c)]. The intensity spectrums of input, output and
compensated pulse are calculated, using discrete Fourier transform
(DFT) and the transmission spectrum,  and are illustrated in Fig.
\ref{FREQUENCY}. It can be seen that the intensity spectrum of the
Gaussian pulse keeps unchanged with the bandwidth of
\mbox{$\sim$74.4 kHz}; by contrast, that of the AMG pulse has
\emph{not only} a resonant Gaussian component, but \emph{also} two
non-resonant Gaussian sidebands with \mbox{$\pm$700 kHz} detunings.
This phenomenon can be qualitatively analyzed from the Fourier
intensity spectrums of two input pulses:
\begin{align}
\mathrm{Gaussian:}\ I_1(\Delta)&=\exp\left[-\frac{(\ln2)\Delta^2}{\Omega_0^2}\right],\nonumber \\
\mathrm{AMG:}\
I_2(\Delta)&=I_1(\Delta)+\frac{1}{4}\left[I_1(\Delta-\delta)+I_1(\Delta+\delta)\right],\nonumber
\end{align}
where  $I_1(\Delta)$ and $I_2(\Delta)$ are the intensities of two
spectrums, respectively; $\Delta$ is the detuning from resonance;
$\Omega_0$ (determined by $T_0$) is the FWHM of the Gaussian pulse
spectrum. Suppose the system is linear and
$E_{out}(\Delta)=A(\Delta)e^{i\Phi{(\Delta)}}E_{in}(\Delta)$, where
$A(\Delta)$ [$\Phi(\Delta)$] is a real function related to the
absorption (dispersion). Note that each spectral component has a
different transmission rate that depends on its detuning. For AMG
pulse, the sideband components undergo much more severe absorption
than the resonant one, and so cause more significant distortion. In
this way, we can explain that it is the frequency dependent
absorption that causes the loss and distortion, rather than ``the
pulses `forget' their initial temporal shapes", as stated in \cite
{Niko2005}.

\begin{figure}[htb]
\centerline{\includegraphics[width=8.0cm]{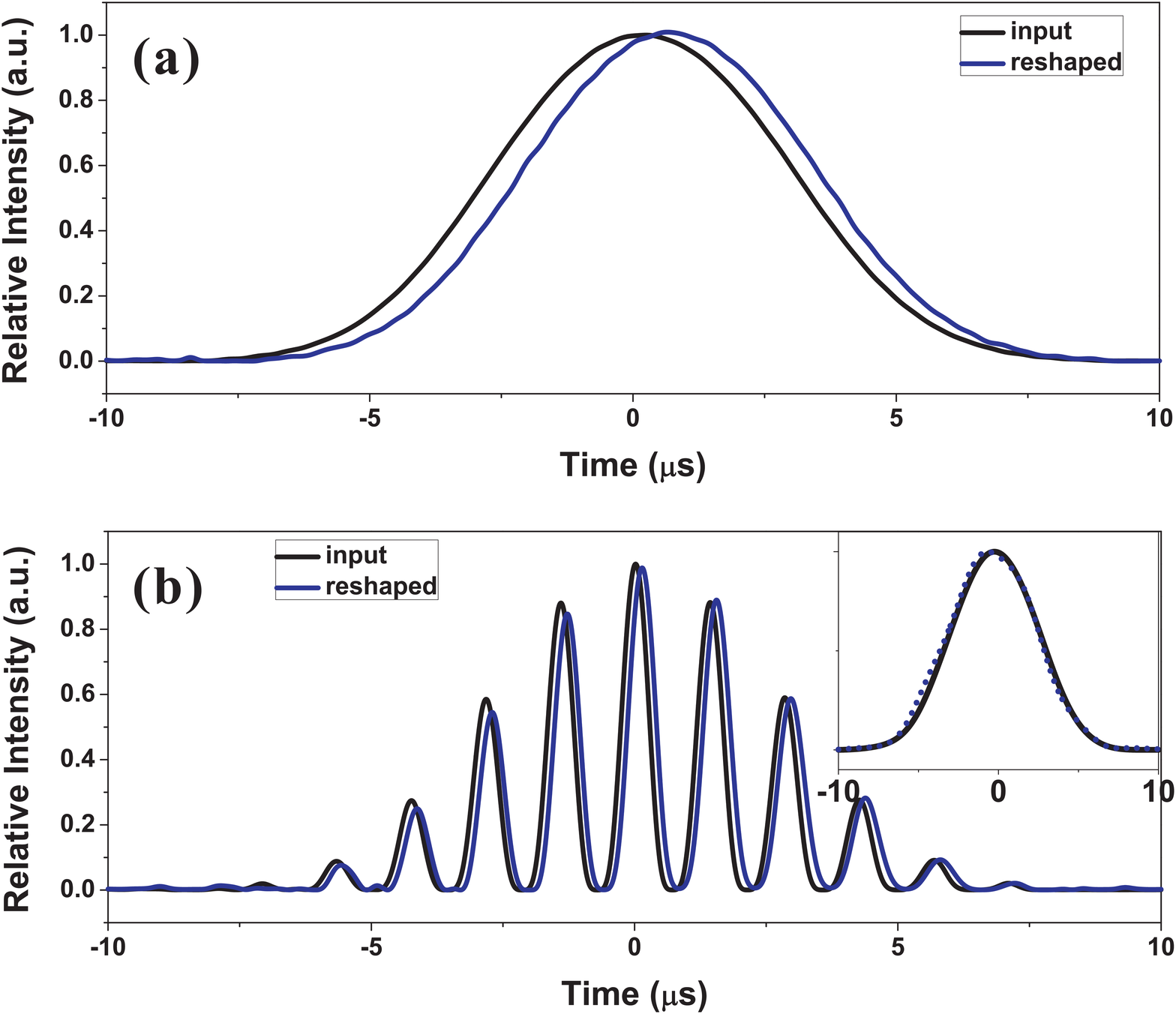}}
\caption{(color online). Input and reshaped (calculated from the
compensated spectrum in Fig. \ref{FREQUENCY}) pulses versus time.
(a) Gaussian. (b) AMG. The inset shows the fast light (obtained by
IDFT) of the two sideband components of the compensated spectrum in
Fig. \ref{FREQUENCY}(b). The reshaped pulse (blue dotted) is
advanced by a little time relative to the corresponding input one
(black solid).}\label{RESHAPED}
\end{figure}

Therefore, for the linear absorption, we have
$I_{out}(\Delta)=|A(\Delta)|^2I_{in}(\Delta)$, where the phase
information is eliminated, and $|A(\Delta)|^2$ is the transmission
spectrum in Fig. \ref{SETUP}(c). A compensation method is thus
proposed and applied to reshape the output pulse. As shown in Fig.
\ref{FREQUENCY}, we obtain the compensated spectrums by simply
amplifying the output spectrums according to the transmission rate
as $I_{comp}(\Delta)=I_{out}(\Delta)/|A(\Delta)|^2_{measured}$.
Since the compensated and the input shown in Fig. \ref{FREQUENCY}
are well overlapped, this method can \emph{recover} the pulse with
high fidelity. To further confirm this, we transform the compensated
spectrum back to time domain using inverse discrete Fourier
transform (IDFT), and plot the input and the reshaped pulses (Fig.
\ref{RESHAPED}). The reshaped pulses of Gaussian and AMG, with
different time delays, are both in good agreement with their
corresponding input ones in shape and intensity. Even with the
significant distortion, the output of AMG pulse has still been
compensated well. Thus, only considering the frequency dependent
absorption is sufficient to compensate for the distortion in a
single-$\Lambda$ type system.

It should be noted that the delay time of the reshaped AMG pulse is
less than that of the Gaussian one, as shown in Fig. \ref{RESHAPED}.
This is because the three separate spectral components of the AMG
pulse experience different dispersions. By transforming each
spectral component back to time domain separately, we obtain three
Gaussian pulses with absolutely different time ``delays". For the
resonant component, we obtain a slow light of a Gaussian pulse with
the same delay time as in Fig. \ref{RESHAPED}(a). On the contrary,
for each sideband component, we obtain a Gaussian pulse with a
little advance [fast light, see the inset of Fig.
\ref{RESHAPED}(b)]. As a result, the time delay for the AMG pulse is
decreased. This effect, know as the simultaneous slow and fast light
\cite{YifuZhu2006}, inevitably causes distortion (``dispersive
distortion"), but it is negligibly small for a single-$\Lambda$ type
system \cite{Boyd2005,Camacho2006}.

In conclusion, the slow light of an amplitude modulated pulse is
studied, according to our knowledge, for the first time
experimentally. We demonstrate that the loss and distortion of the
amplitude modulated pulse are significant, which is not desirable in
application. Using spectrum analysis, we analyze the loss and
distortion in frequency domain in a single-$\Lambda$ type system,
and point out that the distortion is dominantly caused by different
absorption rate for each spectral component. Accordingly, we present
a compensation method to reshape slow light pulse with high
fidelity. Further, the dispersion effect is analyzed, which shows
that though the ``dispersive distortion" is negligible, this effect
reveals a novel way, different from \cite{YifuZhu2006}, to realize
simultaneous slow and fast light. The aforementioned results can be
extended to an arbitrary amplitude modulated pulse in a
single-$\Lambda$ type system in EIT. These techniques of amplitude
modulation and reshaping in slow light have potential applications
in optical buffer and the low-distortion optical delay lines.

This work is supported by the Key Project of the National Natural
Science Foundation of China (Grant No. 60837004).

\end{document}